\documentclass[pra,twocolumn,floatfix,superscriptaddress,longbibliography]{revtex4-1}

\usepackage[T1]{fontenc}
\usepackage{natbib}
\usepackage{graphicx}
\usepackage{indentfirst}
\usepackage{listings}
\usepackage{braket}
\usepackage{amsmath}
\usepackage{amssymb}
\usepackage{amsthm}
\usepackage[font=small,labelfont=bf]{caption}
\usepackage{float}
\usepackage{circuitikz}
\usepackage{caption}
\usepackage{subcaption}
\usepackage{array, makecell}
\usepackage{bbold}
\usepackage{placeins}

\usepackage[title]{appendix}
\newtheorem{definition}{Definition}

\usepackage{qcircuit}

\usepackage{tikz}
\usepackage{lipsum}

\begin{document}

\title{Quantum computation capability verification protocol for NISQ devices with dihedral coset problem}

\author{Ruge Lin}
\affiliation{Technology Innovation Institute, Abu Dhabi, UAE.}
\affiliation{Departament de F\'{i}sica Qu\`{a}ntica i Astrof\'{i}sica and Institut de Ci\`{e}ncies del Cosmos (ICCUB), Universitat de Barcelona, Mart\'{i} i Franqu\`{e}s 1, 08028 Barcelona, Spain.}
%\email{ruge.lin@tii.ae}
\author{Weiqiang Wen}
\affiliation{LTCL, Telecom Paris, Institut Polytechnique de Paris, France.}
%\email{weiqiang.wen@telecom-paris.fr}

\begin{abstract}

In this article, we propose an interactive protocol for one party (the verifier) holding a quantum computer to verify the quantum computation power of another party's (the prover) device via a one-way quantum channel. This protocol is referred to as the dihedral coset problem (DCP) challenge. The verifier needs to prepare quantum states encoding secrets (DCP samples) and send them to the prover. The prover is then tasked with recovering those secrets with a certain accuracy. Numerical simulation demonstrates that this accuracy is sensitive to errors in quantum hardware. Additionally, the DCP challenge serves as benchmarking protocol for locally fully connected (LFC) quantum architecture and aims to be performed on current and near-future quantum resources. We conduct a $4$-qubit experiment on one of the IBM Q devices.

\end{abstract}

\maketitle

\section{Introduction}

In 2019, Google succeeded in reaching quantum supremacy with their Sycamore processor \cite{Supremacy}. However, it remains a long way to a fully functioning quantum computer. At this moment, only noisy intermediate-scale quantum (NISQ) \cite{NISQ} devices are available, and a method is needed to verify their computing power. 

Currently, instead of computation capability, random circuit sampling and cross-entropy benchmarking \cite{RCS,CEB} are primarily concerned with testing the quantum property of the device. It is desirable to have a performance test on quantum hardware, proving to a verifier and unable to falsify. Recent works \cite{Regev09,Brakerski0,Brakerski1,Zhu} demands a classical verifier. In particular, they rely on the hardness of the learning with errors (LWE) problem and needs thousands of qubits, which is not applicable to present quantum hardware.

This test should be designed based on two principles: dynamic enough to adapt various processors and friendly to NISQ devices, which can be directly applied in an experiment. In order to be dynamic, we focus on LFC quantum architecture. LFC means that the chip consists of $m$ unit cells of $n+1$ qubits, with $m \geq 2$ and $n \geq 1$. Within each cell, $n+1$ qubits are fully connected, and each cell has a leader qubit, $m$ leader qubits are fully connected. LFC shares many similarities with Chimera and Pegasus topologies in quantum annealing processor D-Wave \cite{Dwave}. Notice that in reality, hardware for gate-based quantum computing rarely follow this geometry, but $SWAP$ gates can be applied. A test based on LFC structure can cover any quantum chip with the number of qubits $\geq 4$ and not prime. Moreover, a quantum device should pass a test based on LFC architecture to demonstrate its potential for fully connected circuits, such as Shor algorithm \cite{Shor} and Grover algorithm \cite{Grover}. Furthermore, for applying to NISQ devices, the test should contain only shallow circuits and not rely on quantum memory.

Nowadays, classical simulation programs for quantum circuits such as Cirq \cite{Cirq}, Qiskit \cite{Qiskit} and Qibo \cite{Qibo,QiboGithub} can mimic noisy or noiseless quantum devices for up to dozens of qubits on classical hardware. It is hard to distinguish between a quantum device and a simulator around this scale. Therefore, we can consider introducing a quantum verifier. In previous works \cite{fitzsimons2018post,takeuchi2021divide}, the quantum verifier(s) is(are) asked to witness particular states generated by the prover. However, in \cite{fitzsimons2018post}, the target state is too complicated for NISQ devices. Also, the method provided in \cite{takeuchi2021divide} is designed for sparse quantum chips with certain geometry restrictions.

This article presents the DCP challenge, a verification protocol of quantum computation capability, requiring a quantum verifier and a one-way quantum channel from the verifier to the prover. It is an interactive protocol for Alice, the verifier holding a $n+1$-qubit quantum device, to test the quantum computing power of Bob, the prover holding a $m\times\left(n+1\right)$-qubit device, which runs on the LFC architecture. In contrast to the method in \cite{takeuchi2021divide} where the verifier needs more than half of the qubits of the prover, the DCP challenge only needs a fraction, implying a quantum channel with fewer qubits. In particular, Alice needs to provide simple quantum states (DCP samples) as a superposition of two possibilities, which can be easily verified by measurement, and send them to Bob, who solves the problem essentially using Quantum Fourier transform on $n$ qubits. The advantage of the prover being the receiver of the quantum states is that the measurement error is also tested. We have also performed simulations of our protocol. On one side, we show that in the error-free model, the quantum computing capability of the prover can be successfully verified with overwhelming probability. On the other side, in the noisy setting simulation, our protocol is shown to be very sensitive to the presence of errors, while it is still shown to be robust up to some restricted errors. This property also makes the DCP challenge a promising benchmarking protocol when preparing samples and solving the problem are performed by the same quantum device.

\section{Preliminary}

\subsection{Dihedral coset problem}\label{ssec:dcp}

The dihedral coset problem has been a fundamental problem in studying the quantum hardness of the hidden subgroup problem over (non-abelian) dihedral group in the last two decades \cite{MEPH,GSVV01,Regev02,FIMSS03,HRTS00,RoBe98}. Informally, it asks to recover the hidden subgroup of a dihedral group given random cosets of the hidden subgroup as superposition. A dihedral group is generated by reflections and rotations of a $E$-gon (regular polygon with $E$ edges). The first part of the superposition encodes the reflection. From now on, we call it the reflection qubit. The second part encodes the rotation. Normalization is omitted for every equation in this article.

\begin{definition}[Dihedral coset problem, DCP]\label{def:DCP}
The input of the DCP$_{E}^{\ell}$ with modulus $E$ consists of $\ell$ samples. Each sample is a quantum state of the form

\begin{equation}
    \ket{\psi_{x,s}}=\ket{0}\ket{x} \  + \ \ket{1}\ket{\left(x+s\right) \bmod E},
\end{equation}

stored in $1+\lceil\log_2 E\rceil$ qubits, where $x\in\{0,1,...,E-1\}$ is randomly and uniformly selected for each sample and $s\in\{0,1,...,E-1\}$ is fixed throughout all the states. The task is to output the secret $s$. 

\end{definition}

The problem is hypothesized to be unsolvable by direct measurement on the computational basis, which means the best-known classical solution is a random guess. We could not obtain $x$ and $\left(x+s\right) \bmod E$ at the same time.

The DCP is known to be solvable in sub-exponential time while given a sub-exponential number of samples \cite{Kuperberg05,Regev04,Kuperberg13}. These solving algorithms were designed with different optimization targets. So far, Kuperberg's algorithm \cite{Kuperberg05} achieves a smallest running-time $2^{O(\sqrt{\log E})}$ but requires $2^{O(\sqrt{\log E})}$ space while Regev's \cite{Regev04} variant requires only a polynomial (in $\log E$) space but its running-time is slightly worse as $2^{O(\sqrt{\log E\log\log E})}$.

Both of them start by running quantum Fourier transform on the given DCP samples (except the reflection qubit) and measure them, which naturally possess an LFC structure. The main drawback of these two algorithms is that some quantum states need to be maintained throughout the whole process.

In this work, given the constraints of current quantum computing devices (e.g., NISQ), the circuit depth and quantum memory required by both Kuperberg's and Regev's algorithms can not be satisfied.
Therefore, we consider a slightly different variant of the DCP problem and algorithm by minimizing circuit depth and limiting quantum registers. 

Before introducing them, we first recall the quantum Fourier transform.

\begin{definition}[Quantum Fourier transform, QFT]\label{def:QFT}
The quantum Fourier transform on the computational basis $\ket{0},...,\ket{N-1}$ of an $n$ qubit state is defined to be a linear operator with the following action on the basis states,
  
\begin{equation}
    \ket{j} \mapsto \sum_{k=0}^{N-1} \omega_N^{jk} \ket{k},
\end{equation}

where $\omega_N = \textrm{e}^{\frac{2\pi i}{N}}$.
\end{definition}
The evaluation time of QFT is $\mathcal{O}\left(n^2\right)$ \cite[Section~5.1]{NiCh00}.

\subsection{New variant}

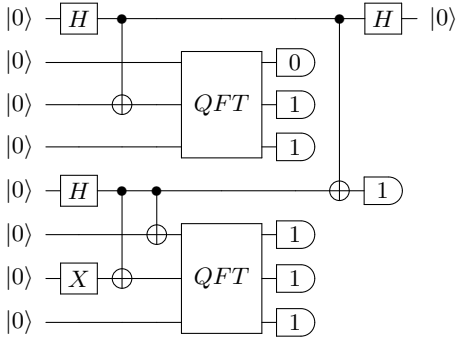
\begin{figure}[t]
\[
\begin{array}{c}
\Qcircuit @C=0.6em @R=0.6em 
{
\lstick{\ket{0}} & \gate{H} & \ctrl{2} & \qw & \qw                & \qw          & \ctrl{4} & \gate{H} & \rstick{\ket{0}} \qw\\
\lstick{\ket{0}} & \qw      & \qw      & \qw & \multigate{2}{QFT} & \measureD{0} \\
\lstick{\ket{0}} & \qw      & \targ    & \qw & \ghost{QFT}        & \measureD{1} \\
\lstick{\ket{0}} & \qw      & \qw      & \qw & \ghost{QFT}        & \measureD{1}\\
\lstick{\ket{0}} & \gate{H} & \ctrl{2} & \ctrl{1} & \qw                & \qw          & \targ    & \measureD{1}\\
\lstick{\ket{0}} & \qw      & \qw      & \targ & \multigate{2}{QFT} & \measureD{1}\\
\lstick{\ket{0}} & \gate{X} & \targ    & \qw & \ghost{QFT}        & \measureD{1}\\
\lstick{\ket{0}} & \qw      & \qw      & \qw & \ghost{QFT}        & \measureD{1}
}
\end{array}
\]
\caption{A toy circuit for $m=2$, $n=3$ and $s=2$, first four qubits correspond to the state when $x=0$, $\ket{\psi_{0,2}}=\ket{0}\ket{000}+\ket{1}\ket{010}$ and the second four qubits correspond to the state $x=2$, $\ket{\psi_{2,2}}=\ket{0}\ket{010}+\ket{1}\ket{100}$. The collision after QFT with $\hat{x}_1=3$ and $\hat{x_2}=7$ is chosen randomly.}
\label{fig: s_even}
\end{figure}

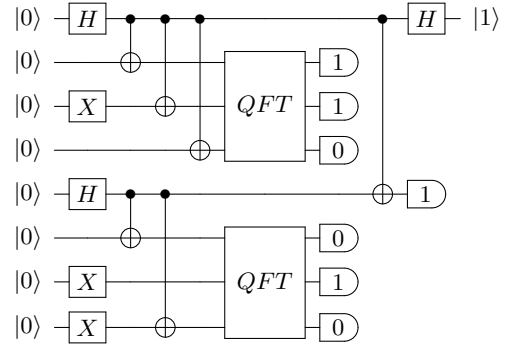
\begin{figure}[t]
\[
\begin{array}{c}
\Qcircuit @C=0.6em @R=0.6em 
{
\lstick{\ket{0}} & \gate{H} & \ctrl{1} & \ctrl{2} & \ctrl{3} & \qw                &\qw           & \ctrl{4}           & \gate{H} & \rstick{\ket{1}} \qw\\
\lstick{\ket{0}} & \qw      & \targ    & \qw      & \qw      & \multigate{2}{QFT} & \measureD{1} \\
\lstick{\ket{0}} & \gate{X} & \qw      & \targ    & \qw      &\ghost{QFT}         & \measureD{1} \\
\lstick{\ket{0}} & \qw      & \qw      & \qw      & \targ    & \ghost{QFT}        & \measureD{0}\\
\lstick{\ket{0}} & \gate{H} & \ctrl{1} & \ctrl{3} & \qw      & \qw                & \qw          & \targ    & \measureD{1}\\
\lstick{\ket{0}} & \qw      & \targ    & \qw      & \qw      & \multigate{2}{QFT} & \measureD{0}\\
\lstick{\ket{0}} & \gate{X} & \qw      & \qw      & \qw      & \ghost{QFT}        & \measureD{1}\\
\lstick{\ket{0}} & \gate{X} & \qw      & \targ    & \qw      & \ghost{QFT}        & \measureD{0}
}
\end{array}
\]
\caption{A toy circuit for $m=2$, $n=3$ and $s=3$, first four qubits correspond to the state when $x=2$, $\ket{\psi_{2,3}}=\ket{0}\ket{010}+\ket{1}\ket{101}$ and the second four qubits correspond to the state $x=3$, $\ket{\psi_{3,3}}=\ket{0}\ket{011}+\ket{1}\ket{110}$. The collision after QFT with $\hat{x}_1=6$ and $\hat{x_2}=2$ is chosen randomly.}
\label{fig: s_odd}
\end{figure}

Currently, NISQ devices have limited registers, low coherence time, low relaxation time, and imperfect gate implementation. They can only efficiently perform shallow circuits. Therefore, we slightly modify the DCP adapting this status. First, we set $E=N=2^{n}$. Then, instead of solving the secret $s$, we ask to solve the parity of $s$, which represents the same order of complexity. FIG. \ref{fig: s_even} and FIG. \ref{fig: s_odd} are two example circuits of this new variant.

Alice can prepare the state $\ket{\psi_{x,s}}$ with only $H$, $X$ and $CNOT$ (which are the Clifford gates) and it takes $\mathcal{O}\left(n\right)$ gates. She can verify the accuracy of $\ket{\psi_{x,s}}$ by measuring it. Notice that for total $N^2$ combinations of $x$ and $s$, there are total $N^2$ combinations of $X$ and $CNOT$ gates. However, we do not have a direct relation between $x$, $s$ and each of these gates.

To solve the parity of $s$ within $m$ cells of $n+1$ qubits, using the shallowest circuit currently known, we use a highly simplified version of Kuperberg's algorithm \cite{Kuperberg05}, and name it \emph{ParitySolve}.

Bob performs QFT on the last $n$ qubits and measures them. Here we highlight that he always needs more computation resources and operation steps than Alice; otherwise, it would not be a challenge.

After QFT is applied on the last $n$ qubits of the DCP sample, the total state becomes

\begin{equation}
    \sum_{k=0}^{N-1}\left(\ket{0}+\omega^{ks}_{N}\ket{1}\right)\ket{k}, \hspace{0.5cm}\omega_N=e^{\frac{2\pi i}{N}}.
\end{equation}

Bob then checks the measurements after QFT. He needs a pair of measurements that the most significant qubit is different and the rest are identical. We call it a collision. If he does not have it, he resets all registers to $\ket{0}$ and starts another \emph{ParitySolve}.

After the measurement, the reflection qubit becomes

\begin{equation}
    \ket{\phi_{\hat{x},s}}=\ket{0}+\omega_N^{\hat{x}s}\ket{1},
\end{equation}

for some uniform distributed random measured $\hat{x}\in\{0,1,...,N-1\}$. Assume that Bob has a collision, $\hat{x}_1$ and $\hat{x}_2$, then the tensor product between $\ket{\phi_{\hat{x}_1,s}}$ and $\ket{\phi_{\hat{x}_2,s}}$ gives

\begin{equation}
\ket{0,0}+\omega_N^{\hat{x}_{1}s}\ket{1,0}+\omega_N^{\hat{x}_{2}s}\ket{0,1}+\omega_N^{\left(\hat{x}_{1}+\hat{x}_{2}\right)s}\ket{1,1}.
\end{equation}

Bob performs a $CNOT$ gate on these two reflection qubits. The state becomes

\begin{equation}
    \ket{0,0}+\omega_N^{\hat{x}_{1}s}\ket{1,1}+\omega_N^{\hat{x}_{2}s}\ket{0,1}+\omega_N^{\left(\hat{x}_{1}+\hat{x}_{2}\right)s}\ket{1,0}.
\end{equation}

Then he measures the target qubits, with $\frac{1}{2}$ probability he can measure $\ket{1}$. If $\ket{0}$ is measured, he needs to reset all registers to $\ket{0}$ and start another \emph{ParitySolve}. After $\ket{1}$ on the target qubit is measured, the controlled qubit becomes

\begin{equation}
    \ket{0}+\omega_N^{\left(\hat{x}_1-\hat{x}_2\right)s}\ket{1}=\ket{0}+\left(-1\right)^{s}\ket{1}.
\end{equation}

The equality holds because if $\hat{x}_1$ and $\hat{x}_2$ is a collision, then $\hat{x}_{1}-\hat{x}_{2} \mod N=\frac{N}{2}$.

Finally, the parity of $s$ lies inside the phase of $\ket{1}$. Bob can solve it by applying an $H$ gate on the remaining qubit and measuring it. If the result is $\ket{0}$, then $s$ is even. He replies $0$ to Alice. Otherwise, $s$ is odd. He replies $1$. The solution is completely correct if the quantum channel and devices are noiseless. 

\subsection{Measurement method}

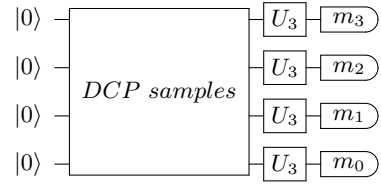
\begin{figure}[t]
\[
\begin{array}{c}
\Qcircuit @C=0.6em @R=0.6em 
{
\lstick{\ket{0}}&\multigate{3}{DCP\hspace{0.1cm}samples}&\gate{U_3}&\measureD{m_3}\\
\lstick{\ket{0}}&\ghost{DCP\hspace{0.1cm}samples}       &\gate{U_3}&\measureD{m_2}\\
\lstick{\ket{0}}&\ghost{DCP\hspace{0.1cm}samples}       &\gate{U_3}&\measureD{m_1}\\
\lstick{\ket{0}}&\ghost{DCP\hspace{0.1cm}samples}       &\gate{U_3}&\measureD{m_0}
}
\end{array}
\]
\caption{A toy circuit for the measurement method for $n=3$.}
\label{fig: measurement circuit}
\end{figure}

We have found a new method to solve the parity of $s$ when $E=N$. As indicated previously, $s$ is unsolvable by direct measurement on the computational basis. However, it is possible to have a minor advantage with single-qubit measurement on a different basis, equivalent to applying one layer of same single-qubit unitary gates before measuring on the computational basis, as shown in FIG. \ref{fig: measurement circuit}.

The third general unitary gate $U_3$ can be written into

\begin{equation}
\begin{aligned}
&U_3\left(a,b,c\right)=
\begin{pmatrix}
e^{-i\left(b+c\right)/2}\cos\left(\frac{a}{2}\right) &  
-e^{-i\left(b-c\right)/2}\sin\left(\frac{a}{2}\right) \\
e^{i\left(b-c\right)/2}\sin\left(\frac{a}{2}\right) & 
e^{i\left(b+c\right)/2}\cos\left(\frac{a}{2}\right)
\end{pmatrix},
\end{aligned}
\end{equation}

with $a\in \left[0,\pi\right)$, $b\in \left[0,4\pi\right)$ and $c\in \left[0,2\pi\right)$. When $a=\frac{\pi}{2}$, $b=0$ and $c=\pi$, $U_3$ is an $H$ gate (with a global phase).

The parity of $s$ can be distinguished with $a=\frac{\pi}{2}$, $c\in\{0,\pi\}$, and an arbitrary $b$. The measurement is read as $M=m_0 2^0 + m_1 2^1 + ... + m_{n} 2^{n}$. We note $M_{non}$ the value not able to measure, $M_{even}$ the value that is only measurable when $s$ is even, $M_{odd}$ the value that is only measurable when $s$ is odd. Therefore, $M_{even}$ and $M_{odd}$ can be considered as the feature values of the parity of $s$, which can be determined when one of them appears. When $c=0$, $M_{non}=N-1$, $M_{even}=2N-2$ and $M_{odd}=N-2$. When $c=\pi$, $M_{non}=N$, $M_{even}=1$ and $M_{odd}=N+1$. The probability of measuring $M_{even}$ or $M_{odd}$ is $\frac{1}{N}$. For example, Bob can measure every DCP samples on an $H$ basis, if any of them is $1$, $s$ is even, or if any of them is $N+1$, then $s$ is odd. This new technique is found by brute-force simulation for $n<10$ and conjectured to be valid for lager $n$, potentially leads to a solution of the DCP with only measurement.

\section{Protocol}

\begin{figure*}[t]
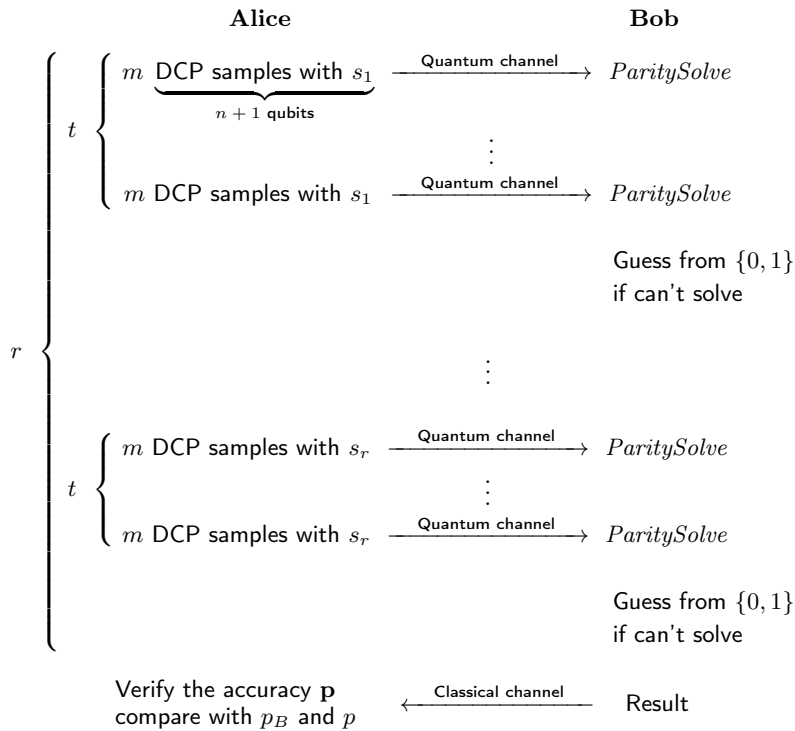

  \hspace{4cm}\textbf{Alice}\hspace{4.5cm}\textbf{Bob}\hspace{3cm}
  \centering
  \[
  \begin{array}{c c}
  r &
  \begin{cases}
  \begin{array}{c c}
  t &
  \begin{cases}
  \begin{array}{c c c}
  m\hspace{0.1cm}\underbrace{\textsf{DCP samples with $s_1$}}_{\textsf{$n+1$ qubits}} & \xrightarrow{\hspace{1em}\textsf{Quantum channel}\hspace{1em}} & \emph{ParitySolve}\\
  & \vdots & \\
  \textsf{$m$ DCP samples with $s_1$} & \xrightarrow{\hspace{1em}\textsf{Quantum channel}\hspace{1em}} & \emph{ParitySolve}
  \end{array}
  \end{cases}
  \end{array}\\
  \\
  \hspace{7.35cm}\textsf{Guess from $\{0,1\}$}\\
  \hspace{7.35cm}\textsf{if can't solve}\\
  \\
  \hspace{5.65cm}\vdots\\
  \\
  \begin{array}{c c}
  t &
  \begin{cases}
  \begin{array}{c c c}
  \textsf{$m$ DCP samples with $s_r$} & \xrightarrow{\hspace{1em}\textsf{Quantum channel}\hspace{1em}} & \emph{ParitySolve}\\
  & \vdots & \\
  \textsf{$m$ DCP samples with $s_r$} & \xrightarrow{\hspace{1em}\textsf{Quantum channel}\hspace{1em}} & \emph{ParitySolve}\\
  \end{array}
  \end{cases}
  \end{array}\\
  \\
  \hspace{7.35cm}\textsf{Guess from $\{0,1\}$}\\
  \hspace{7.35cm}\textsf{if can't solve}\\
  \end{cases}
  \end{array}
  \]
  \hspace{1.6cm}
  \begin{tabular}{rl}
      &\textsf{Verify the accuracy $\mathbf{p}$}\\
      &\textsf{compare with $p_B$ and $p$}
  \end{tabular}
  \hspace{0.3cm} $\xleftarrow{\hspace{1em}\textsf{Classical channel}\hspace{1em}}$ \hspace{0.3cm}\textsf{Result}\hspace{2.3cm}
  \caption{The DCP challenge in a diagram.}
  \label{fig: protocol}
\end{figure*}

In this section, we use an example to demonstrate the full protocol of the DCP challenge. A diagram is in FIG. \ref{fig: protocol}. Alice is the verifier who has a quantum computer. Bob is the prover who declares having a quantum computer and wants to prove his quantum computation capability to Alice. To perform the challenge, Alice needs to have $n+1$ qubits to prepare DCP samples, and Bob needs $m\times\left(n+1\right)$ qubits to solve them. Before starting the challenge, they agree on the choice of $m$, $n$, the number of iterations $t$ and the number of repetitions $r$. In every repetition, there are $t$ iterations.

At the first stage, Alice uniformly selects two numbers $x\in\{0,1,..., N-1\}$ and $s\in\{0,1,..., N-1\}$, which she keeps both of them as secret. She generates $\ket{\psi_{x,s}}$ with $x$ and $s$. Alice sends $m$ DCP samples one by one to Bob via a quantum channel in every iteration. Bob stores them into his $m$ cells of registers and attempts to solve the parity of $s$ using \emph{ParitySolve}. In the first repetition, every sample has the same $s=s_1$ and a different random $x$. When Bob could not have a result after $t$ iterations, he randomly guesses a $0$ or $1$. Here we have $\ell=m\times t$. Then Alice starts new repetitions, each time with a different $s$ until she completes the challenge with a secret $s_r$. The number of repetitions $r$ can be any number large enough to reflect Bob's probability of success, also called the accuracy $\mathbf{p}$. In this article, we use bold letter for simulation or experimental outcome. Bob sends his results in a bit-string back to Alice via a classical channel.

Finally, Alice verifies Bob's probability of success $\mathbf{p}$. If Bob has an error-free device, his accuracy is expected to be $p$, which can be calculated or simulated numerically. Furthermore, the choice of $m$, $n$, $t$ depends on the number of qubits from both parties, the maximum transmission of the quantum channel and the difference $p-p_B$, where $p_B$ is the expected accuracy of performing the measurement method. Details are shown in appendix \ref{app_birthday}. Moreover, the presence of noise also implies the loss of computing power will reflect in the accuracy. Since NISQ hardware is not likely to be error-free, we expect to have $p \geq \mathbf{p} \geq \frac{1}{2}$. The quantum computation capability of Bob's processor is verified with $\mathbf{p}>p_B$. When $\mathbf{p}\sim p$, the device is qualified for a stricter test.

The numerical simulation of this verification protocol can be found in appendix \ref{app_simulation_verification}.

\section{Possible cheating methods}\label{cheating}

There is not any known method to cheat the DCP challenge without a quantum computer of better performance unless a new algorithm is found, reaching the expected accuracy $p$ with a shallower circuit than \emph{ParitySolve}. However, it is possible to cheat when having such a device and obtain $\mathbf{p}>p$ with even less than $m\times\left(n+1\right)$ qubits. Two methods are outlined below.

The first method assumes Bob's quantum computer has a longer relaxation time, such that those unmeasured qubits do not quickly return to $\ket{0}$. Instead of receiving $m$ samples and erasing them all if he could not find a collision, he can erase one sample each time and receive another one. Once a collision is found, and after $CNOT$ gate he measures $\ket{0}$, he can erase both of them and receive another two samples. This method wastes fewer DCP samples and leads to a larger probability of success.

The second method assumes Bob's quantum computer has less noise to perform $SWAP$ gates efficiently. Bob can move reflection qubits to the register of measured qubits after resetting them to $\ket{0}$; therefore, he has more room to store the reflection qubit of every sample. This method increases the probability of collision, thus, increasing the accuracy.

Both methods rely on a more enhanced quantum computation capability, so they should not be considered cheating. Once quantum computers become powerful enough to "cheat" accurately, the "cheating method" can become the standard protocol. All Alice needs to do is to reduce $t$ or raise $p$ accordingly. There are more methods to reach $\mathbf{p}>p$ when Regev's \cite{Regev02,Regev04} and Kuperberg's \cite{Kuperberg05,Kuperberg13} complete algorithms can be performed.

Nonetheless, we can prohibit all these cheating methods by setting the time interval between iterations long enough to bypass the possible relaxation time for the near future but still short enough for an experiment. For example, we can set the interval as one second (Sycamore processor is in the order of $\mu s$), so the device loses all its memory of the previous iteration. In this way, Bob has no cheating method unless he has a quantum computer with an extremely longer relaxation time.

\section{Other applications}

The DCP challenge has more applications than a verification mechanism of quantum computation power. Here are some examples.

This protocol can be used to benchmark a quantum computer. It is a straightforward method for evaluating the performance of NISQ hardware. Gate based quantum devices are usually manufactured using various techniques thus have distinct connection geometry and parameters. Even when they have the same number of qubits, direct comparison of their computation capability is difficult. The readout of the DCP challenge is the numerical accuracy $\mathbf{p}$ after applying a large amount of simple pre-defined circuits. It provides us a quantitative insight into a quantum computer, which can be regarded as a score. The numerical simulation of using the DCP challenge as a benchmarking protocol is in appendix \ref{app_simulation_benchmarking}. The smallest instance using the DCP challenge for benchmarking only requires $4$ qubits in a line. In this case, $QFT$ is an $H$ gate. We perform this experiment on the first $4$ qubits of $5$-qubit IBM Q processor \emph{ibmq\_manila} \cite{IBM}, as detailed in appendix \ref{app_IBM}.

The DCP challenge helps benchmark a quantum channel. If Bob tests on his processor and has a probability of success $\mathbf{p}$. They should anticipate a comparable level of accuracy when Alice transmits the challenge to Bob. This protocol can also help to spot eavesdropping on a quantum channel. If Alice and Bob used to have a probability of success $\mathbf{p}$, suddenly the probability has dropped. If they are both certain there are no technical issues with the channel or their hardware, then perhaps Eve is intercepting. She steals some DCP samples from the channel, and when she puts some fake samples back, she is unaware of the parity of $s$. Even if Eve can also intercept the classical channel from Bob to Alice and change the result, she has no method to raise the probability unless she replaces Bob completely. 

The DCP challenge is a very elemental puzzle game for NISQ devices. Its numerous potentials remain unexplored.

\section{Generalization}

Here we give a more general interactive verification protocol. The central assumption is a question encoding a secret in $\ell$ quantum states (samples), a quantum algorithm solves the secret with a probability $p$, a classical or measurement algorithm solves the secret with a baseline probability $p_B$. The key to verifying the quantum computing power lies in the inequality $p>p_B$. Alice sends a fixed amount $\ell$ of samples in each repetition. Bob needs to solve the secret and sends his result back to Alice, and she verifies the accuracy and compares it with $p_B$. By increasing the number of repetitions, Alice can confirm Bob's quantum computation capability. The protocol can be optimized by lowering the number of qubits and simplifying Alice's process of preparing the samples, creating a computation imbalance between the verifier and prover.

The DCP is chosen with the extra advantage that its current solutions naturally process an LFC structure. Moreover, even within the DCP framework, our readers are free to design new protocols for more advanced quantum computers, for example, Alice can ask about the full $s$ instead of its parity or she can decide another $E<N$. New algorithm for solving the DCP or its different variants will be invented in the future and the protocol will be updated accordingly. However, the sub-exponential quantum complexity of the DCP remains relatively solid since it secures the hardness of LWE problem \cite{brakerski2018learning}.

\section{Conclusions}

In this article, the DCP challenge has been proposed. Its computation has been shown, numerical simulations have been done and different cheating strategies have been evaluated. Other applications have been described and a generalization has been produced. Our readers may perceive it as a quantum game rather than a methodology for confirming quantum processing capacity. Its rules are flexible and can be adapted to different situations.

The DCP challenge is designed for NISQ devices, and it is aimed to serve temporary. One day, when quantum computers are powerful enough to outperform classical computers in various tasks such as factoring big integers, the protocol will lose its purpose as a proof of computation. Nevertheless, its other applications remain.

\begin{acknowledgements}

We acknowledge Stavros Efthymiou and Sergi Ramos-Calderer for useful support in Qibo, Ingo Roth for useful suggestion. Also, we acknowledge the use of IBM Quantum services for this work.

\end{acknowledgements}

%\bibliographystyle{unsrt}
%\bibliography{mainbib}

\appendix

\section{Analytical probability}\label{app_birthday}

To obtain the ideal probability of success $p$, we need to determine $k_{collision}$, the probability of NOT having a collision in $m$ cells among $N$ possibilities. We can use the formula of "Birthday Paradox" to provide its upper bound and a lower bound.

$k_{lower}$ is a direct application of the formula to calculate the probability of NOT having two identical elements when choosing $m$ times out of $N$ possibilities,

\begin{equation}
    k_{lower}=\prod_{i=0}^{m-1} \frac{N-i}{N}.
\end{equation}

In order NOT to have a pair of identical elements, each choice must be different. However, in the case of NOT having a collision, we can keep the same choice. So it is slightly easier to have a collision than to have a pair of identical elements.

For calculating $k_{upper}$, we first consider the probability of NOT having two identical elements when choosing $m$ times out of $\frac{N}{2}$ possibilities (for $n-1$ qubits except the most significant one). Then, we take into account that the last qubit is different. We have

\begin{equation}
    k_{upper}=\frac{1}{2}+\frac{1}{2}\prod_{i=0}^{m-1} \frac{N/2-i}{N/2}.
\end{equation}

$k_{upper}$ ignores the case of having more than one pair of identical elements when we are considering the first $n-1$ qubits.

We have

\begin{equation}
    k_{upper} > k_{collision} > k_{lower}.
\end{equation}

But each collision has only a probability of $\frac{1}{2}$ to solve the parity of $s$, so the chance of NOT being able to solve after $t$ iterations is

\begin{equation}
    2-2p=\left(\frac{1+k_{collision}}{2}\right)^t.
\end{equation}

Finally, when Bob is unable to solve, he has to randomly guess a result, which has $\frac{1}{2}$ to be correct. So in total, his probability of success is

\begin{equation}
    p=\left(2-\left(\frac{1+k_{collision}}{2}\right)^t\right)/2.
    \label{eq_p}
\end{equation}

And we have the relation,

\begin{equation}
    p_{upper} > p > p_{lower},
\end{equation}

with

\begin{equation}
    p_{upper}=\left(2-\left(\frac{1+k_{lower}}{2}\right)^t\right)/2,
    \label{eq_p_upper}
\end{equation}

and

\begin{equation}
    p_{lower}=\left(2-\left(\frac{1+k_{upper}}{2}\right)^t\right)/2.
    \label{eq_p_lower}
\end{equation}

Although we do not have the analytical expression of $p$, we can obtain it numerically. We can generate $m$ random bit strings of length $n$, search for collision. By repeating this procedure, we can have the numerical $k_{collision}$ and use it for calculating $p$.

In order to have a given $p$, we can have an estimation of $t$,

\begin{equation}
    t\sim\frac{\log\left(2-2p\right)}{\log\left(\frac{1+k_{lower}}{2}\right)}.
    \label{eq_t}
\end{equation}

We use $k_{lower}$ instead of $k_{upper}$ because it is numerically closer to $k_{collision}$. The choice of $t$ is flexible, but setting it too large is not just a waste of resources: if Bob can solve the parity of $s$ multiple times within $t$ iterations and take the majority result, the protocol becomes less sensitive to error. 

Here we consider only the case when measuring $\ket{0}$ after $CNOT$; we reset all registers and pass directly into the next iteration, ignoring the fact that there might be another collision in the same group. This is because the probability of having two collisions in the same group is significantly lower than having one, and this difference vanishes in $p$ with the exponent $t$. We would like to keep the problem as simple as possible. Also, we would like to maintain the shallowest circuit. This situation is easy to simulate classically.

In some situations, especially when $N \gg m$, $t$ can be too large ($>1,000$) to fit in an experiment. In this case, we can set a lower $p$ and increase $r$ for preciseness. The rules of this protocol are adjustable. For a numerical indication, for a device with $1,000$ qubits, if we set $n=19$ and $m=50$, $t$ should be $\geq 785$ to have $p>80\%$ according to our protocol. Therefore, our protocol is still feasible for the advanced NISQ-era. At that time, quantum computers might be capable of "cheating" accurately (as in section \ref{cheating}), we can even set a lower $t$.

There is a probability of $2-2p$ that a random guess needs to be made. Using the standard deviation formula, we can calculate the fluctuation from the expectation $p$,

\begin{equation}
    \sigma_p=\sqrt{\frac{1-p}{2r}}.
    \label{eq_SD}
\end{equation}

This formula of standard deviation is also valid for experimental $\mathbf{p}$. When $\mathbf{p}=\frac{1}{2}$, the fluctuation becomes the same as flipping a coin $r$ times.

For the measurement method, the probability of measuring $M_{even}$ or $M_{odd}$ is $\frac{1}{N}$ and the probability of measuring both of them in the same repetition is $0$. Therefore, the probability of measuring none of them, also means not able to solve the parity of $s$ within $m\times t$ samples is

\begin{equation}
    2-2p_B=\left(\frac{N-1}{N}\right)^{mt}.
\end{equation}

The expected accuracy is

\begin{equation}
    p_B=\left(2-\left(\frac{N-1}{N}\right)^{mt}\right)/2.
    \label{eq_p_B}
\end{equation}

The standard deviation can also be calculated with the Eq. (\ref{eq_SD}).

\begin{figure*}[t]
\begin{subfigure}{0.32\linewidth}
\centering
\includegraphics[width=\textwidth]{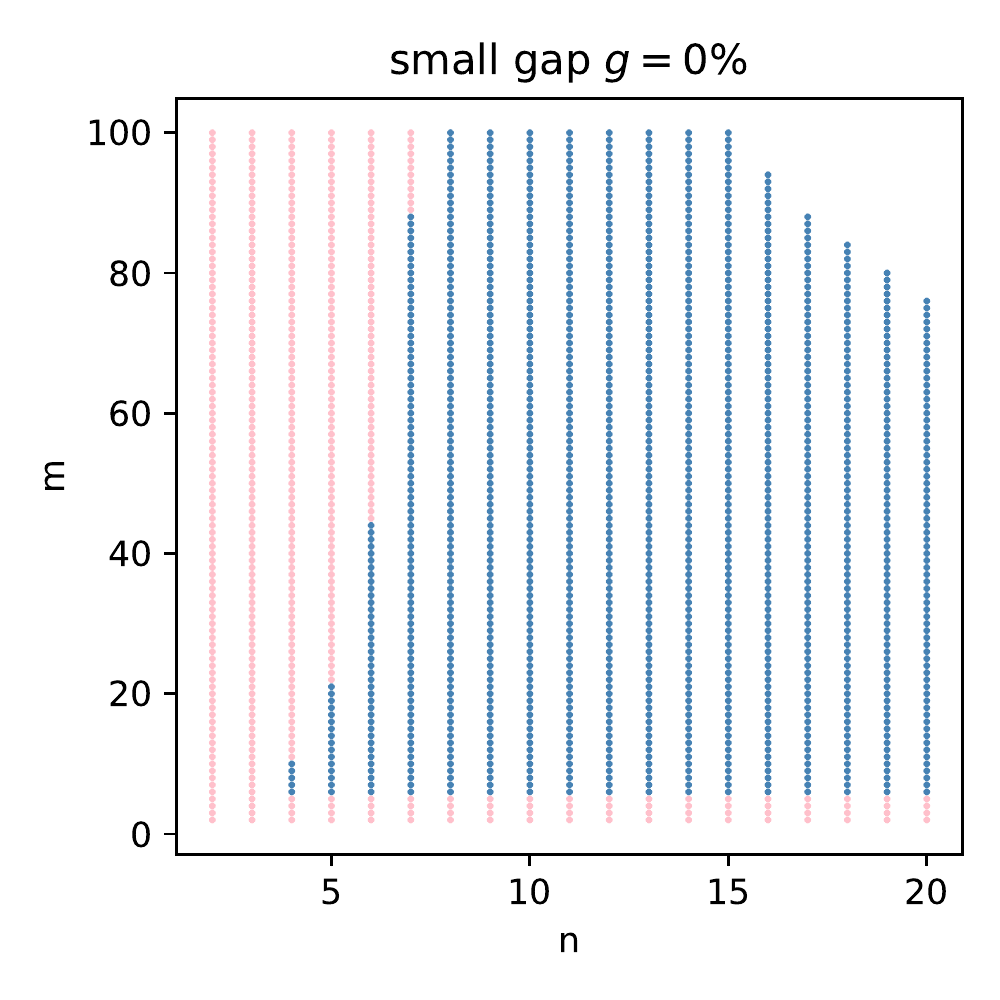}
\caption{Smallest instance: $m=6$, $n=4$, $t=1$ with $p_{B}=66.05\%$ and $p_{upper}=66.40\%$.}
\label{fig: small_gap}
\end{subfigure}
\begin{subfigure}{0.32\linewidth}
\centering
\includegraphics[width=\textwidth]{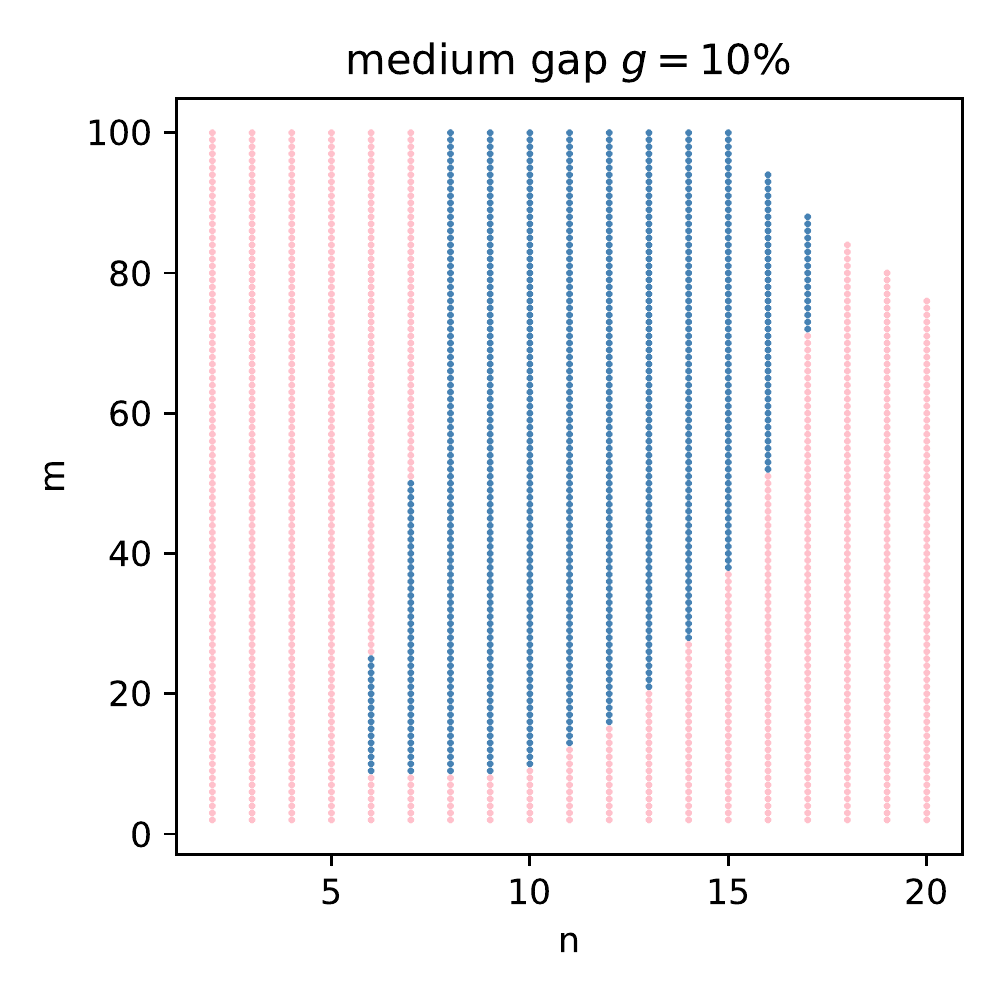}
\caption{Smallest instance: $m=9$, $n=6$, $t=4$ with $p_{B}=71.64\%$ and $p_{upper}=81.73\%$.}
\label{fig: mid_gap}
\end{subfigure}
\begin{subfigure}{0.32\linewidth}
\centering
\includegraphics[width=\textwidth]{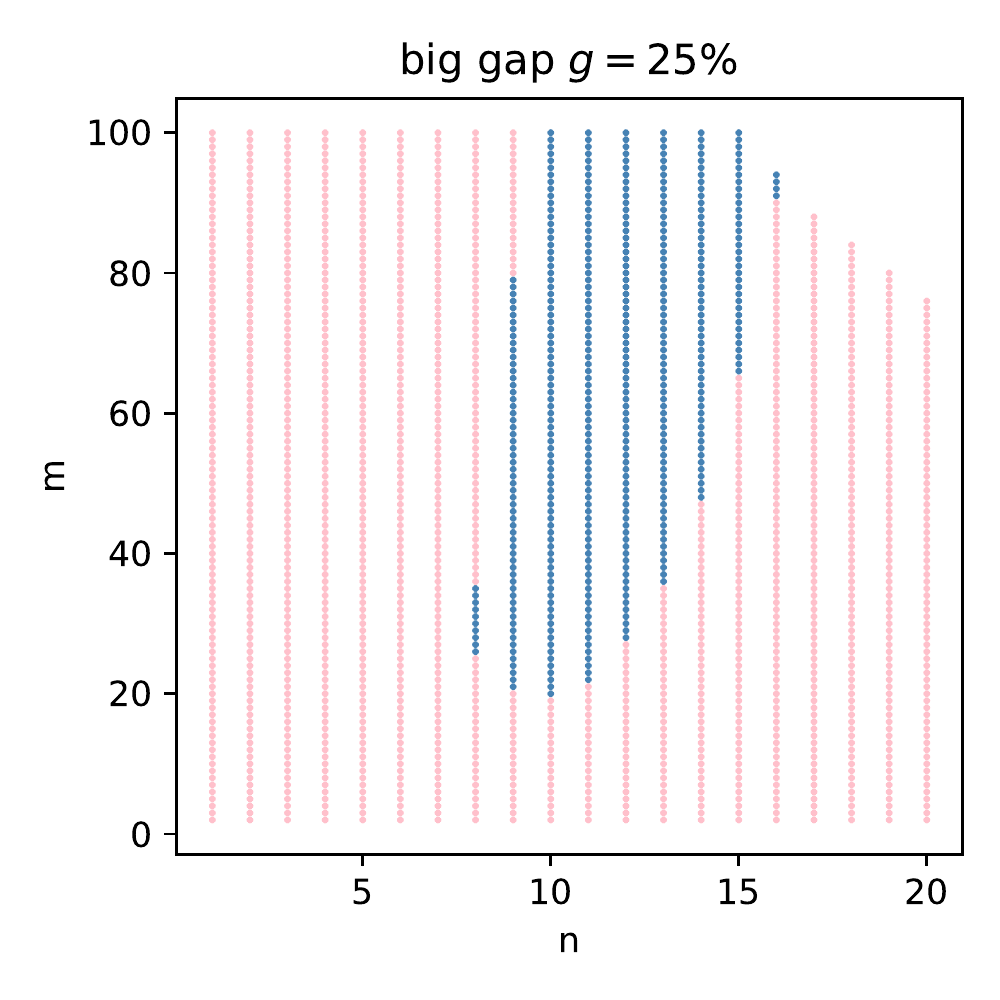}
\caption{Smallest instance: $m=21$, $n=9$, $t=9$ with $p_{B}=65.45\%$ and $p_{upper}=90.66\%$.}
\label{fig: big_gap}
\end{subfigure}
\caption{Combinations of $m$ and $n$ that we can have $p_{upper}-p_B>g$ with a $t<25$ are plotted in blue (darker). These three figures show the situation of small gap $g=0\%$, medium gap $g=10\%$ and big gap $g=25\%$ for less than $1,600$ qubits.}
\label{fig:three graphs}
\end{figure*}

From the Eq. (\ref{eq_p_upper}) and the Eq. (\ref{eq_p_B}), we can compare $p_{upper}$, which is numerically closer to $p$, and $p_{B}$ for difference $m$, $n$ and $t$. For less than $1,600$ qubits, a minor difference is shown in FIG. \ref{fig: small_gap} for an indication. The number $1600$ is the most up-to-date lower bound of a verification protocol with a classical verifier \cite{Zhu}, ideally a quantum verifier is no longer need after this scale. A distinguishable difference is shown in FIG. \ref{fig: mid_gap}, \emph{ParitySolve} and the measurement method can be distinguished for $r\sim 1,000$. A significant difference is shown in FIG. \ref{fig: big_gap}, in this case, the quantum computation capability of Bob can be verified even with moderated error.

\section{Numerical simulation for verification}\label{app_simulation_verification}

One advantage of the DCP challenge is that it is effortless to simulate the whole process classically. Due to the LFC structure and the shallowness of the circuit, instead of simulating the entire circuit of $m\times\left(n+1\right)$ qubits, we can simulate each DCP sample individually and store the measurement bit-string and the state vector of the remaining qubit. Qibo can efficiently simulate a quantum circuit for up to $31$ qubits on a laptop. Therefore, it can simulate a DCP circuit for up to $n=30$ qubits.

\begin{figure}[t]
  \centering
  \includegraphics[width=1\linewidth]{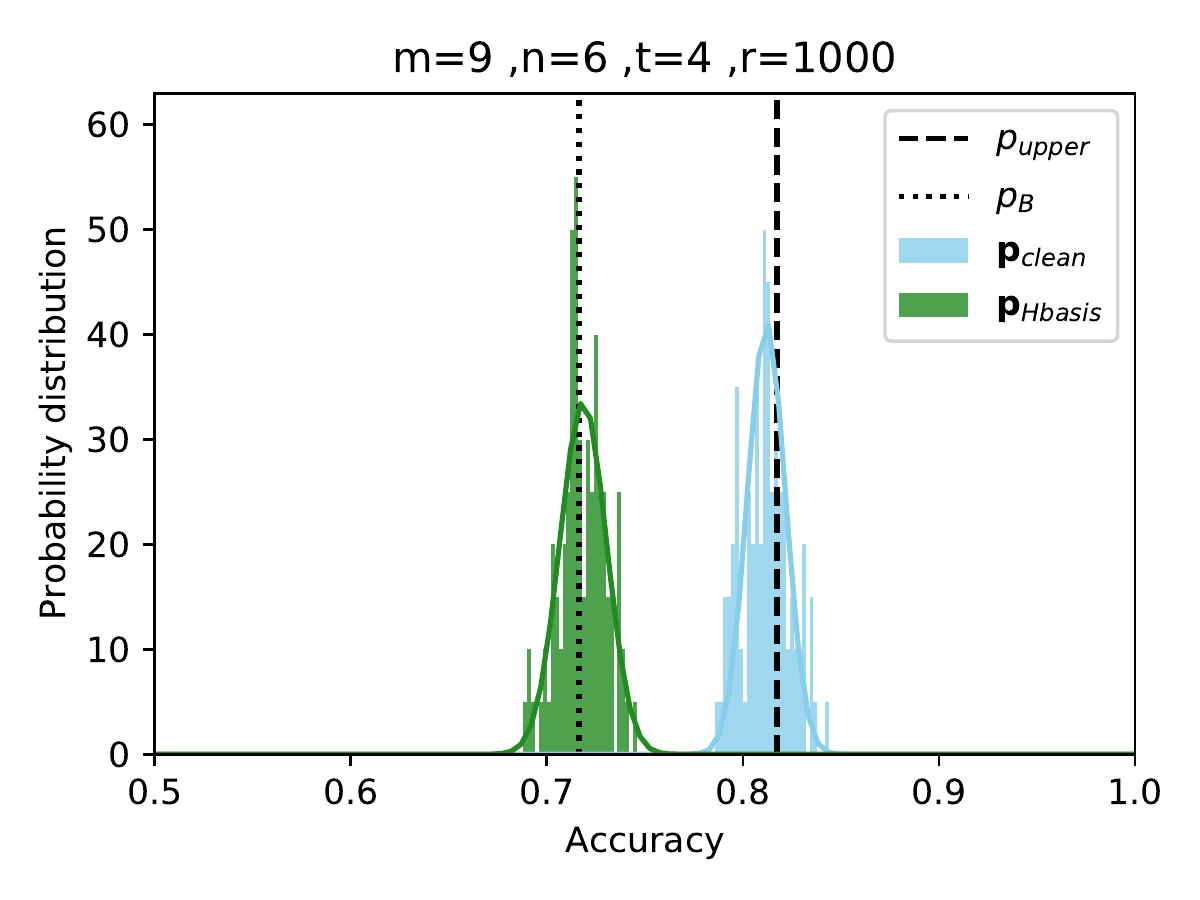}
  \caption{Normalized probability distribution of $\mathbf{p}_{clean}$ and $\mathbf{p}_{Hbasis}$ over $100$ trials, plotted with normal distribution $\mathcal{N}\left(\overline{\mathbf{p}_{clean}},\sigma_{\mathbf{p}_{clean}}\right)$ and $\mathcal{N}\left(\overline{\mathbf{p}_{Hbasis}},\sigma_{\mathbf{p}_{Hbasis}}\right)$.}
  \label{fig: verification}
\end{figure}

If Bob has a quantum chip of $m=9$, and $n=6$, Alice can first choose with FIG. \ref{fig: mid_gap} that $t=4$. Then she can calculate $p_{upper}$ with Eq. (\ref{eq_p_upper}) and $p_{B}$ with Eq. (\ref{eq_p_B}), or even calculate $p$ using Eq. (\ref{eq_p}) with a numerical $k_{collision}$, to see the probability that she expects. She prepares $r=1,000$ repetitions to challenge Bob, each with $m=9$ samples of $n+1=7$ qubits, and $t=4$ iterations. In total, she needs to prepare $36,000$ DCP samples, and Bob needs to perform about that many QFTs. Finally, Bob sends his $1,000$ answers back to Alice, verifying his accuracy. FIG. \ref{fig: verification} is a simulation with Qibo of $\mathbf{p}_{clean}$ the accuracy of error-free simulation of \emph{ParitySolve} and $\mathbf{p}_{Hbasis}$ the accuracy of solving by measuring on the $H$ basis. We have $\overline{\mathbf{p}_{clean}}=p$ and $\overline{\mathbf{p}_{Hbasis}}=p_B$. The code is on Github \cite{Github}. We consider $r=1,000$ acceptable since it is trivial to distinguish the probability distribution of the clean circuit performing \emph{ParitySolve} and the measurement method despite fluctuation.

\section{Numerical simulation for benchmarking}\label{app_simulation_benchmarking}

We use $p\sim 80\%$ for simulation. It is not too close to $50\%$, so we can notice the decrease of accuracy due to noise. Our readers can also choose $p\sim 90\%$, which means $t$ needs to be $\sim 1.75$ times greater according to Eq. (\ref{eq_t}). TABLE \ref{simulation} is the table of accuracy, a comparison between analytical $p_{lower}$, $p_{upper}$, error-free circuit simulation $\mathbf{p}_{clean}$ and noisy circuit simulation $\mathbf{p}_{error}$ \cite{Github}. For the noise map, we use $1\%$ for bit error and phase error, $3\%$ for measurement error, the choice of errors is inspired by Quantum Computer Datasheet \cite{datasheet} from Google.

In the table, $p_{upper}>\mathbf{p}_{clean}>p_{lower}$ for $m<8$, which is what we expected. But $p$ can be very close to $p_{upper}$. As we can see, when $m=8$, $\mathbf{p}_{clean}$ is actually larger than $p_{upper}$ because we obtain it through sampling and there is minor fluctuation.

\begin{figure}[t]
  \centering
  \includegraphics[width=1\linewidth]{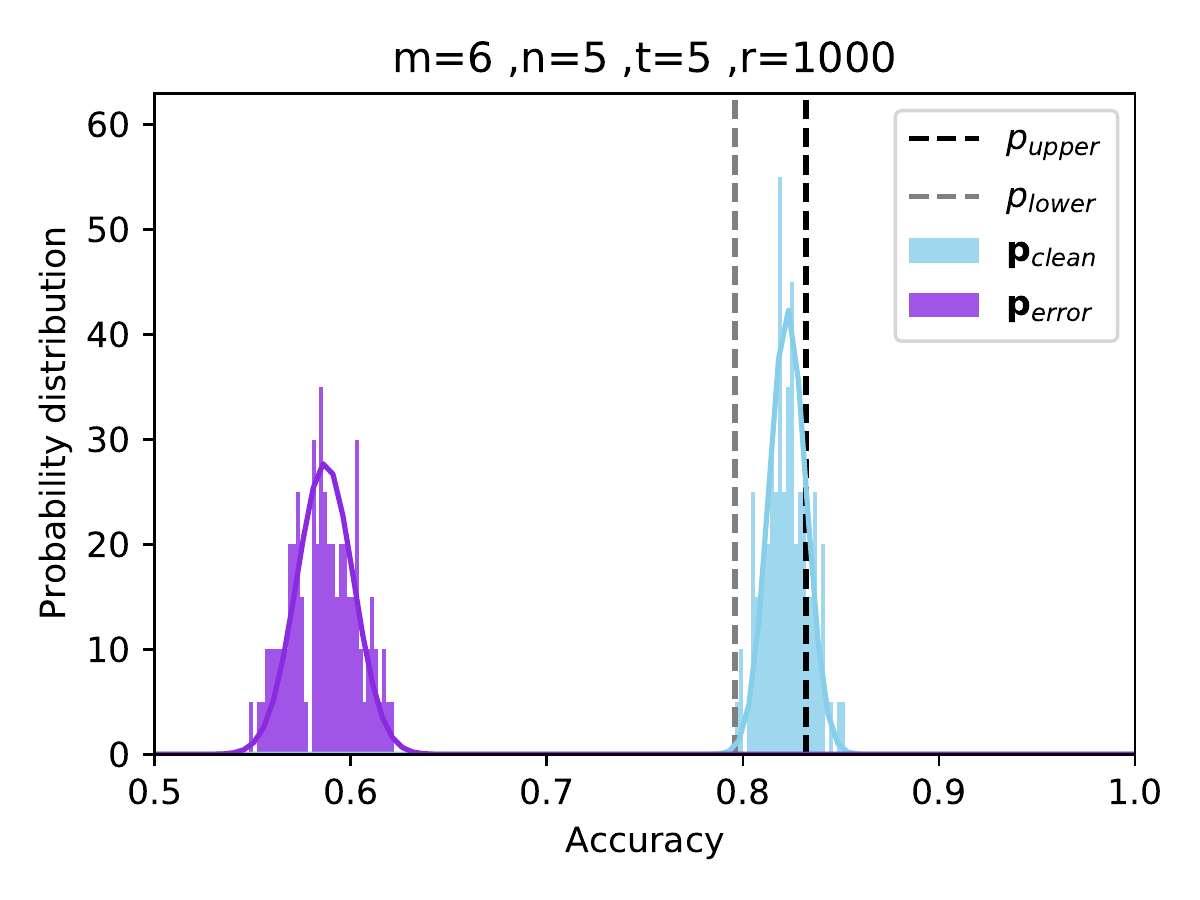}
  \caption{Normalized probability distribution of $\mathbf{p}_{clean}$ and $\mathbf{p}_{error}$ over $100$ trials, plotted with normal distribution $\mathcal{N}\left(\overline{\mathbf{p}_{clean}},\sigma_{\mathbf{p}_{clean}}\right)$ and $\mathcal{N}\left(\overline{\mathbf{p}_{error}},\sigma_{\mathbf{p}_{error}}\right)$.}
  \label{fig: PD}
\end{figure}

To benchmark a quantum chip with $n=5$ and $m=6$, we can first predict with Eq. (\ref{eq_t}) that $t=5$, then calculate $p_{upper}$ with Eq. (\ref{eq_p_upper}) and $p_{lower}$ with Eq. (\ref{eq_p_lower}), or even calculate $p$ using Eq. (\ref{eq_p}) with a numerical $k_{collision}$, to see the probability that we expect. If we set $r=1,000$, we need to prepare $30,000$ DCP samples in total and perform about that many QFTs. FIG. \ref{fig: PD} is a simulation of $\mathbf{p}_{clean}$ and $\mathbf{p}_{error}$ in this case. Comparing to $\mathbf{p}_{clean}$, the accuracy $\mathbf{p}_{error}$ shifts towards $\frac{1}{2}$ due to the noise. We consider $r=1,000$ acceptable since it is trivial to distinguish the probability distribution of the clean circuit and the noisy circuit despite fluctuation.

\begin{table}[t]
\centering
\begin{tabular}{ |c|c|c|c|c|c|c| } 
 \hline
 $m$ & 3 & 4 & 5 & 6 & 7 & 8\\ 
 \hline
 $t$ & 3 & 3 & 4 & 5 & 7 & 9\\
 \hline
 $p_{upper}$ & 83.75\% & 82.47\% & 84.18\% & 83.22\% & 83.17\% & 80.62\%\\ 
 \hline
 $p_{lower}$ & 78.90\% & 76.86\% & 79.38\% & 79.59\% & 80.61\% & 78.91\%\\ 
 \hline
 $\mathbf{p}_{clean}$& 81.60\% & 80.86\% & 83.05\% & 82.45\% & 82.38\% & 80.69\%\\
 \hline
 $\mathbf{p}_{error}$& 69.02\% & 65.38\% & 61.77\% & 58.74\% & 55.65\% & 53.30\%\\
 \hline
\end{tabular}
\caption{Here we assume $m=n+1$, so there are in total $m^2$ qubits, and $t$ is the minimal number of iterations for $p_{upper}>80\%$. $\mathbf{p}_{clean}$ and $\mathbf{p}_{error}$ are results of $r=10,000$.}
\label{simulation}
\end{table}

\section{IBM Q experiment}\label{app_IBM}

We benchmark the DCP challenge on $4$ superconducting qubits provided by IBM Q quantum computers. Since the IBM Quantum Composer interface does not allow applying gates after measurement, the DCP challenge can not be directly implemented. We need to perform multiple experiments on every possible configuration then use the output data to reconstruct the DCP challenge.

\begin{figure}[t]
%\centering
\begin{subfigure}[t]{0.4\linewidth}
\centering
\[
\begin{array}{c}
\Qcircuit @C=0.35em @R=0.35em 
{
\lstick{\ket{0}} & \qw      & \gate{H} & \qw      & \measureD{q_0}\\
\lstick{\ket{0}} & \gate{H} & \ctrl{1} & \gate{H} & \measureD{q_1}\\
\lstick{\ket{0}} & \gate{H} & \targ    & \qw      & \measureD{q_2}\\
\lstick{\ket{0}} & \qw      & \gate{H} & \qw      & \measureD{q_3}
}
\end{array}
\]
\caption{Case A, with $s=0$, $x_0=0$ and $x_1=0$.}
\label{fig: caseA}
\end{subfigure}
\begin{subfigure}[t]{0.4\linewidth}
\centering
\[
\begin{array}{c}
\Qcircuit @C=0.35em @R=0.35em 
{
\lstick{\ket{0}} & \qw      & \gate{H} & \qw      & \measureD{q_0}\\
\lstick{\ket{0}} & \gate{H} & \ctrl{1} & \gate{H} & \measureD{q_1}\\
\lstick{\ket{0}} & \gate{H} & \targ    & \qw      & \measureD{q_2}\\
\lstick{\ket{0}} & \gate{X} & \gate{H} & \qw      & \measureD{q_3}
}
\end{array}
\]
\caption{Case B, with $s=0$, $x_0=0$ and $x_1=1$.}
\label{fig: caseB}
\end{subfigure}
\\
\begin{subfigure}[t]{0.4\linewidth}
\centering
\[
\begin{array}{c}
\Qcircuit @C=0.35em @R=0.35em 
{
\lstick{\ket{0}} & \gate{X} & \gate{H} & \qw      & \measureD{q_0}\\
\lstick{\ket{0}} & \gate{H} & \ctrl{1} & \gate{H} & \measureD{q_1}\\
\lstick{\ket{0}} & \gate{H} & \targ    & \qw      & \measureD{q_2}\\
\lstick{\ket{0}} & \qw      & \gate{H} & \qw      & \measureD{q_3}
}
\end{array}
\]
\caption{Case C, with $s=0$, $x_0=1$ and $x_1=0$.}
\label{fig: caseC}
\end{subfigure}
\begin{subfigure}[t]{0.4\linewidth}
\centering
\[
\begin{array}{c}
\Qcircuit @C=0.35em @R=0.35em 
{
\lstick{\ket{0}} & \gate{X} & \gate{H} & \qw      & \measureD{q_0}\\
\lstick{\ket{0}} & \gate{H} & \ctrl{1} & \gate{H} & \measureD{q_1}\\
\lstick{\ket{0}} & \gate{H} & \targ    & \qw      & \measureD{q_2}\\
\lstick{\ket{0}} & \gate{X} & \gate{H} & \qw      & \measureD{q_3}
}
\end{array}
\]
\caption{Case D, with $s=0$, $x_0=1$ and $x_1=1$.}
\label{fig: caseD}
\end{subfigure}
\\
\begin{subfigure}[t]{0.4\linewidth}
\centering
\[
\begin{array}{c}
\Qcircuit @C=0.35em @R=0.35em 
{
\lstick{\ket{0}} & \qw      & \targ     & \gate{H} & \qw      & \measureD{q_0}\\
\lstick{\ket{0}} & \gate{H} & \ctrl{-1} & \ctrl{1} & \gate{H} & \measureD{q_1}\\
\lstick{\ket{0}} & \gate{H} & \ctrl{1}  & \targ    & \qw      & \measureD{q_2}\\
\lstick{\ket{0}} & \qw      & \targ     & \gate{H} & \qw      & \measureD{q_3}
}
\end{array}
\]
\caption{Case E, with $s=1$, $x_0=0$ and $x_1=0$.}
\label{fig: caseE}
\end{subfigure}
\begin{subfigure}[t]{0.4\linewidth}
\centering
\[
\begin{array}{c}
\Qcircuit @C=0.35em @R=0.35em 
{
\lstick{\ket{0}} & \qw      & \targ     & \gate{H} & \qw      & \measureD{q_0}\\
\lstick{\ket{0}} & \gate{H} & \ctrl{-1} & \ctrl{1} & \gate{H} & \measureD{q_1}\\
\lstick{\ket{0}} & \gate{H} & \ctrl{1}  & \targ    & \qw      & \measureD{q_2}\\
\lstick{\ket{0}} & \gate{X} & \targ     & \gate{H} & \qw      & \measureD{q_3}
}
\end{array}
\]
\caption{Case F, with $s=1$, $x_0=0$ and $x_1=1$.}
\label{fig: caseF}
\end{subfigure}
\\
\begin{subfigure}[t]{0.4\linewidth}
\centering
\[
\begin{array}{c}
\Qcircuit @C=0.35em @R=0.35em 
{
\lstick{\ket{0}} & \gate{X} & \targ     & \gate{H} & \qw      & \measureD{q_0}\\
\lstick{\ket{0}} & \gate{H} & \ctrl{-1} & \ctrl{1} & \gate{H} & \measureD{q_1}\\
\lstick{\ket{0}} & \gate{H} & \ctrl{1}  & \targ    & \qw      & \measureD{q_2}\\
\lstick{\ket{0}} & \qw      & \targ     & \gate{H} & \qw      & \measureD{q_3}
}
\end{array}
\]
\caption{Case G, with $s=1$, $x_0=1$ and $x_1=0$.}
\label{fig: caseG}
\end{subfigure}
\begin{subfigure}[t]{0.4\linewidth}
\centering
\[
\begin{array}{c}
\Qcircuit @C=0.35em @R=0.35em 
{
\lstick{\ket{0}} & \gate{X} & \targ     & \gate{H} & \qw      & \measureD{q_0}\\
\lstick{\ket{0}} & \gate{H} & \ctrl{-1} & \ctrl{1} & \gate{H} & \measureD{q_1}\\
\lstick{\ket{0}} & \gate{H} & \ctrl{1}  & \targ    & \qw      & \measureD{q_2}\\
\lstick{\ket{0}} & \gate{X} & \targ     & \gate{H} & \qw      & \measureD{q_3}
}
\end{array}
\]
\caption{Case H, with $s=1$, $x_0=1$ and $x_1=1$.}
\label{fig: caseH}
\end{subfigure}
\caption{Eight possible cases for two DCP samples of $n=1$.}
\label{cases}
\end{figure}
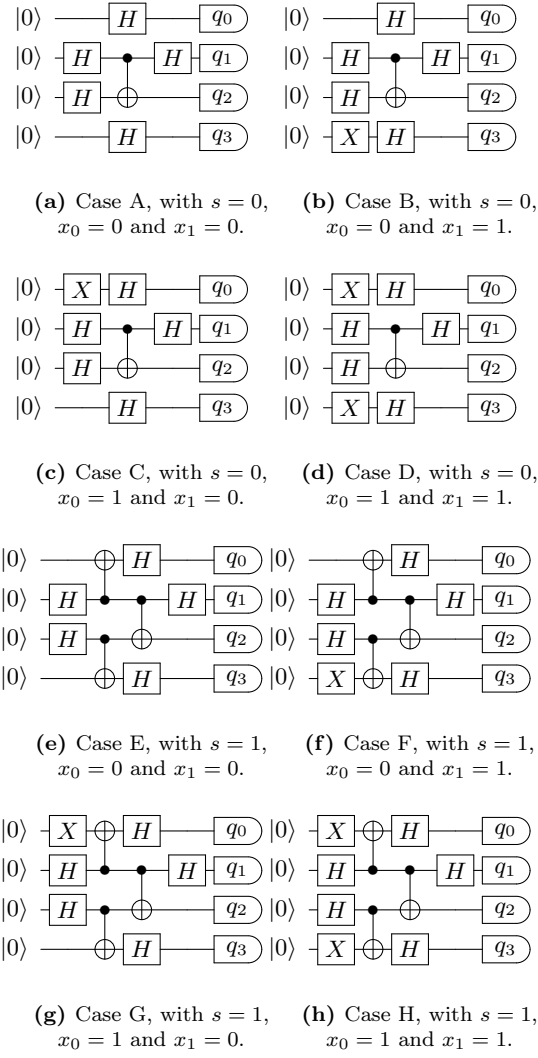

When $n=1$ and $m=2$, there are in total $2^3=8$ possible cases for two DCP samples, as shown in FIG. \ref{cases}. Notice that the reflection qubits are in the centre to avoid $SWAP$ gates. We perform five tests of each case on the first $4$ qubits of $5$-qubit quantum processor \emph{ibmq\_manila}, which has a linear architecture. By default, each test consists of $1024$ shots. Then we select the measurements that have a collision and the result is $\ket{1}$ on the target qubit of $CNOT$ gate, $q_0 \neq q_3$ and $q_2=1$. If the device is noiseless, we should have $q_1=s$.

The result is in TABLE. \ref{IBM}. We can see that the error when $s=1$ is more significant since the circuit has more $CNOT$ gates for preparing DCP samples. The difference will decrease with larger $n$. Eventually, the essential gate-error will be on the $QFT$s or $SWAP$ gates depending on the structure of the device.

\vspace{1cm}

Furthermore, due to the imbalanced measurement error \cite{datasheet}, it is more likely to measure $\ket{0}$ than $\ket{1}$ on a current quantum processor. The same situation can also be caused by the low relaxation time. If Alice chooses $s$ uniformly, Bob is very likely to have more $0$ than $1$ in his result, and he can have a rough estimation of his performance. However, this extra information does not allow him to cheat since he does not know which $0$ should be replaced by $1$.

\begin{figure}[H]
\centering
\includegraphics[width=1\linewidth]{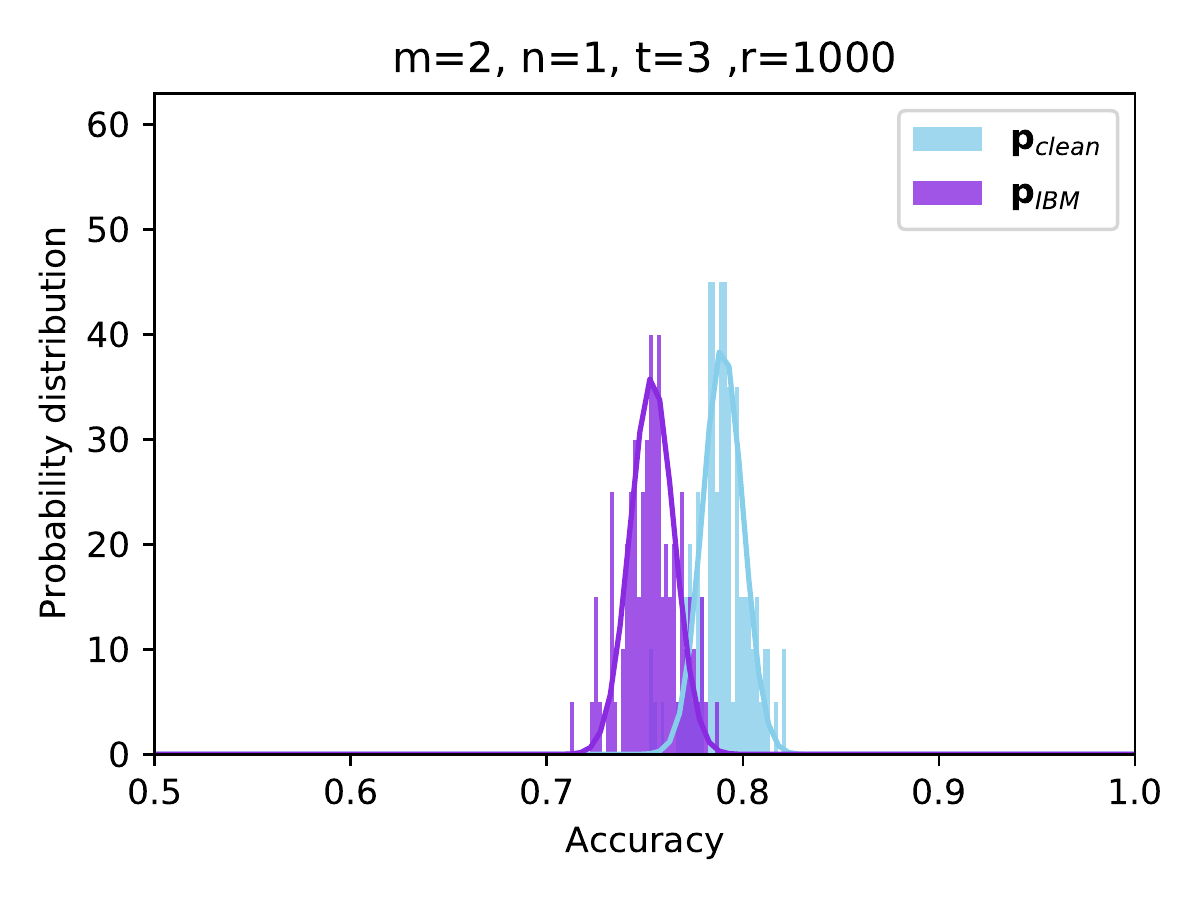}
\caption{Normalized probability distribution of $\mathbf{p}_{clean}$ and $\mathbf{p}_{IBM}$ (reconstructed from experimental data) over $100$ trials, plotted with normal distribution $\mathcal{N}\left(\overline{\mathbf{p}_{clean}},\sigma_{\mathbf{p}_{clean}}\right)$ and $\mathcal{N}\left(\overline{\mathbf{p}_{IBM}},\sigma_{\mathbf{p}_{IBM}}\right)$. The accuracy $\mathbf{p}_{IBM}$ is very close to $\mathbf{p}_{clean}$. A larger $r$ is needed to distinguish them. The quantum device is considered promising.}
\label{fig: IBM}
\end{figure}

We use the erroneous data to reconstruct the DCP challenge \cite{Github}, the result is shown in FIG. \ref{fig: IBM}. The performance of \emph{ibmq\_manila} is not perfect but satisfying. Our reader can also use the DCP challenge to benchmark other processors of IBM Q, such as \emph{ibmq\_santiago}.

\begin{table}[t]
\centering
\begin{tabular}{ |c|c|c|c|c|c|c|c|c| } 
 \hline
 $q_3 q_2 q_1 q_0$ & A & B & C & D & E & F & G & H\\ 
 \hline
 $\ket{0101}$ & 517 & 638 & 624 & 642 & 89   & 73   & 61  & 78   \\
 \hline
 $\ket{0111}$ & 9   & 9   & 14  & 17  & 603  & 563  &526  & 560  \\ 
 \hline
 $\ket{1100}$ & 682 & 575 & 632 & 583 & 56   & 50   &51   & 75   \\ 
 \hline
 $\ket{1110}$ & 20  & 14  & 13  & 13  & 477  & 490  &552  & 545  \\
 \hline
 error        &2.4\%&1.9\%&2.1\%&2.4\%&11.8\%&10.5\%&9.4\%&12.2\%\\
 \hline
\end{tabular}
\caption{Post-selected measurements from IBM Q \emph{ibmq\_manila} processor. In total there are $5426$ shots measuring $q_1=0$ and $4425$ shots measuring $q_1=1$. There are $>1,000$ shots per each cases, which allows us to reconstruct the DCP challenge with $r=1,000$.}
\label{IBM}
\end{table}

\end{document}